\documentclass[conference, 9pt]{IEEEtran}
\IEEEoverridecommandlockouts
\usepackage{cite}
\usepackage{amsmath,amssymb,amsfonts}
\usepackage{algorithmic}
\usepackage{graphicx}
\usepackage{textcomp}
\usepackage{colortbl}
\usepackage{highlight}
\usepackage[hidelinks]{hyperref}
\usepackage{booktabs}
\usepackage{array}
\usepackage{enumitem}
\usepackage{multirow}
\usepackage{adjustbox}
\usepackage{subcaption}

\newcolumntype{C}[1]{>{\centering\arraybackslash}m{#1}}
\def\BibTeX{{\rm B\kern-.05em{\sc i\kern-.025em b}\kern-.08em
    T\kern-.1667em\lower.7ex\hbox{E}\kern-.125emX}}
\begin{document}

\title{Scalable Controllable Accented TTS}

\author{
\IEEEauthorblockN{{Henry Li Xinyuan, Zexin Cai, Ashi Garg, Kevin Duh,}}
\IEEEauthorblockN{{Leibny Paola Garc\'ia-Perera, Sanjeev Khudanpur, Nicholas Andrews, Matthew Wiesner}}
\IEEEauthorblockA{Human Language Technology Center of Excellence
\\Johns Hopkins University
\\Baltimore, United States
\\xli257@jhu.edu
}
}

\maketitle

\begin{abstract}
We propose a method to scale accented TTS training to large, accent-diverse datasets that often lack consistent, high-quality accent labels. Our approach relies on a speech geolocation model to infer accent labels directly from audio. To improve speaker generalization and encourage disentangling speaker from accent we explore timbre augmentation through kNN voice conversion. We validate our approach on CommonVoice by fine-tuning XTTS-v2 with accent labels inferred or improved via geolocation. According to various automated metrics based on embeddings extracted from an accent identification model, the resulting accented TTS model produces speech with better accent fidelity compared to XTTS-v2 fine-tuned on self-reported accent labels in CommonVoice, or other existing accented TTS models. According to human evaluation, it was clear that the geolocation model based data discovery and enhancement improved the naturalness and accent fidelity of generated speech. However, the effect of different data augmentation strategies was less clear.
\end{abstract}


\section{Introduction}
State-of-the-art (SOTA) text-to-speech (TTS) and Voice Conversion systems are increasingly designed to support zero-shot, speech-conditioned synthesis that  enables them to mimic the timbre and speaking style of a reference utterance~\cite{cosyvoice2, maskgct, xtts, genvc, livespeech, valle2, nansy}. However, excplicit control over other speech characteristics, such as prosody, emotion, and especially accent, remains challenging.

A crucial bottleneck in training accented speech synthesis models is the scarcity of large, accent-annotated TTS datasets. Existing TTS datasets that include diverse accents with reliable labels, such as CMU Arctic~\cite{cmuarctic}, L2-Arctic~\cite{l2arctic}, and VCTK~\cite{vctk}, have total durations that only range between 10 and 50 hours.

TTS systems tend to produce American and British accents ~\cite{mascst, accentbox, adv}, the accents most well-represented in commonly used TTS datasets. This bias overlooks the needs of a broader global audience, many of whom speak English with diverse regional accents.
 
While several models for accented TTS have been trained on these datasets and demonstrate good accent similarity~\cite{cvae, mlvae, adv, dart}, they struggle to generalize to unseen speakers due to the limited number of distinct speakers per accent in these datasets.
The high costs of collecting and annotating accent labels severely limit the scalability of these datasets and large, in-the-wild, accent-diverse, speech remains hard to leverage for training controllable accented TTS systems. 

An alternative to manually curated accent-labeled datasets is the use of in-the-wild datasets with diverse accents, one notable example being CommonVoice~\cite{commonvoice}, a crowd-sourced speech dataset with self-reported accent labels. These labels were used to construct the accent-annotated speech corpus CommonAccent~\cite{commonaccent}, an accent-annotated corpus that has enabled accented TTS models such as AccentBox~\cite{accentbox}.
However, CommonAccent still has several limitations:

\begin{enumerate}[leftmargin=*, itemsep=0em]
    \item Low-quality accent labels, particularly due to L2 (non-native) speakers self-reporting as speakers of mainstream accents such as American or Southeastern English accents;
    \item Limited applicability to datasets without self-reported accent labels, such as web-crawled corpora;
    \item Certain accents are represented by very few speakers, resulting in difficulties disentangling speaker timbre from accent characteristics.
\end{enumerate}

\subsection{Towards Scalability in Controllable Accented TTS}

In this work, we introduce Scalable Controllable Accented TTS. To address the challenge of limited scalable accent data, we leverage a speech geolocation model~\cite{geolocating}---a system trained to predict the location on Earth where each utterance was spoken. By using such a model, we can automatically generate accent labels on speech data without accent labels, or enhance the quality of self-reported accent labels. 
Unlike classifier-based accent labeling models such as GenAID~\cite{accentbox}, this method can be extended to any accent in any language without existing labeled speech.
We demonstrate that the precision of accent labels inferred from the geolocation model is similar to SOTA accent-ID systems.

Furthermore, in order to address the lack of speaker diversity for certain accents, we promote speaker-accent separation by using kNN-VC~\cite{knn_vc} to convert each utterance to a diverse array of speaker timbres. We show that kNN-VC preserves the accent of the original utterance, making it well-suited for data augmentation for accented TTS training. 
To validate our proposed method for scaling accent-labeled speech data, we fine-tune a pre-trained multilingual TTS model, XTTS-v2~\cite{xtts}, for accented TTS.

Our contributions are as follows:

\begin{enumerate}[leftmargin=*, itemsep=0em]
    \item We apply the speech geolocation model on accent label discovery and filtering, finding that label filtering using the geolocation model during training improves the quality of accent synthesis, while accent label discovery allows for accented TTS training with no pre-existing labels.
    \item We find that applying kNN-VC as a data augmentation method for training accented TTS systems improves the quality of synthetic speech according to objective metrics.
    \item We achieve comparable or stronger performance to SOTA systems on the accented TTS task on a wide range of accents through XTTS-v2 fine-tuning.
\end{enumerate}

Sample utterances synthesized using our system are included in our demo\footnote{\url{https://hstehstehste.github.io/Projects/Demo/index.html}}.

\section{Related Work}

One way around the lack of labeled accented speech is to leverage the well-established research on accented phonetic transliteration. Since many of the TTS models developed during the years 2020 and 2023 contained an explicit phonemizer module, they could easily be cascaded with a dialectal phonemizer to produce accented speech~\cite{accented_tts_with_limited_data, robinson2023, joint_sequence_model_interpolation, explicit_intensity_control, fine_and_coarse}. The idea has persisted in modern end-to-end TTS models that abandonned the phonemizer, but use large language models or other transliteration models to created for grapheme-level accent transliteration~\cite{mascst, phoneme_level_bert}.

Another approach to address the lack of accented speech data in accents influenced by an L1 substrate language, is to employ zero-shot adaptation using a TTS model trained on the L1 substrate ~\cite{without_parallel_data, radmmm}. The relative scarcity of accented speech data can also be addressed with data augmentation~\cite{accentspeech, without_frontend}. Alternatively, training a TTS model from scratch without sufficient accented speech can be avoided with multi-stage training, where a TTS system is first pre-trained on clean read speech in English or US accent and then fine-tuned on accented speech~\cite{accentspeech}.

Specific modeling strategies are employed to promote speaker-accent disentanglement when the training data is insufficient. The most common strategy is to employ a low-bandwidth accent bottleneck representation, followed by a VAE-based structure which attempts to reconstruct accented speech from the low-bandwidth bottleneck~\cite{accentspeech, dart, cvae, accent_vits, gst}. Another strategy to disentangle speaker and accent representations is domain adversarial training ~\cite{accentbox, multi_scale_modeling, adv}. Data collection efforts have filled in some of the gaps within accented speech datasets. These include datasets that target specific regions of the world~\cite{1000_african}, or that aim to cover the widest possible a range of accents~\cite{globe}.

\section{Method}

\subsection{Automatic Accent Label Discovery and Label Filtering through Speech Geolocation} \label{appproaches}

\begin{figure}[htbp]
\centerline{\includegraphics[width=0.5\textwidth]{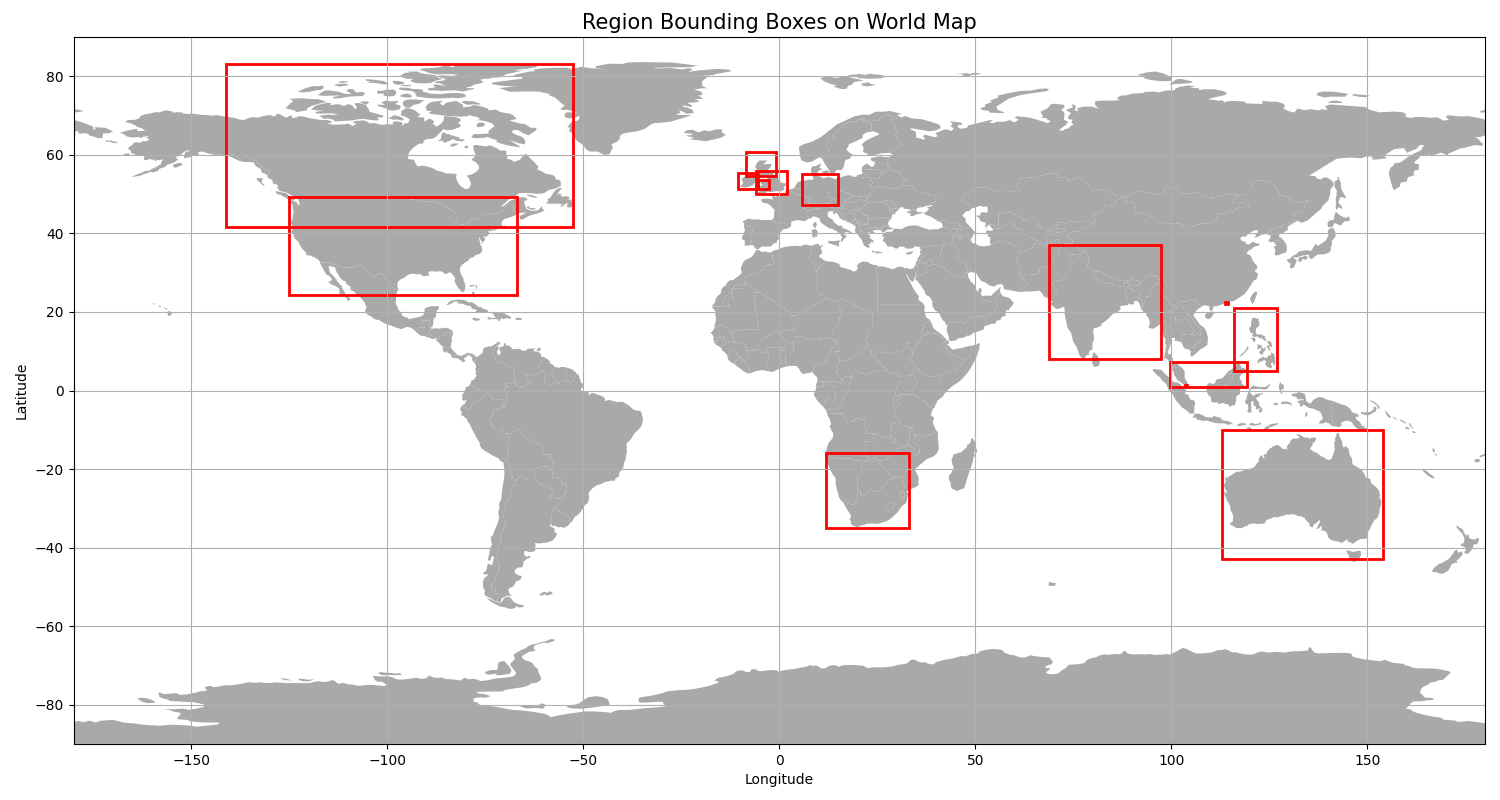}}
\caption{Bounding boxes for each of the accents included in this study.}
\label{world}
\end{figure}

\cite{geolocating} introduced a speech geolocation model trained to predict the broadcast location of speech clips from radio stations around the world. We noticed that this model, though originally developed for applications to language ID, appears, in the process, to learn something about dialect region. In our work, we extend this approach to uncover and refine English accent labels. Specifically, we define approximate geographic bounding boxes for each accent under study (see Figure~\ref{world}). Next, we predict the location of each utterance in CommonVoice using the geolocation model (in a zero-shot fashion, \textbf{without any additional model training}). We \textbf{accept}  utterances as belonging to a target accent if their predicted location falls within the corresponding bounding box.

We source training and testing data from version 20 of CommonVoice~\cite{commonvoice}, which contains a large but potentially noisy collection of accented speech data. To construct accent-specific subsets of CommonVoice for TTS fine-tuning, we compare the following data selection strategies:

\begin{enumerate}[leftmargin=-0.5pt, itemsep=0em] 
    \item \textbf{Unfiltered}: using self-reported accented labels in CommonVoice without any filtering. This is the same approach that was used to construct CommonAccent~\cite{commonaccent}. \label{unfiltered_approach}
    \item \textbf{Filtered}: filtering self-reported accent labels using the geolocation model. \label{filtered_approach}
    \item \textbf{Unlabeled}: discovery of accent labels using the geolocation model. 
\end{enumerate}

\begin{table}[htb]
\caption{Accent Data Statistics in CommonVoice}
\begin{center}
\begin{tabular}{|c||c|c|c||c|}
\toprule
\multicolumn{1}{|c||}{} & \multicolumn{3}{c||}{\textbf{Amount of Data} (hr)} & \multicolumn{1}{c|}{} \\
\midrule
\textbf{Accent} & Labeled & Filtered & Found & $\uparrow$ Found Prec. \\
\hline
US & 432.5 & 222.7 & 511.9 & 72.8$\%$ \\
England & 133.8 & 35.2 & 29.5 & 52.2$\%$ \\
India & 130.7 & 113.1 & 196.4 & 88.8$\%$ \\
Canada & 90.4 & 64.8 & 701.4 & 13.3$\%$ \\
Australia & 67.5 & 34.8 & 68.9 & 80.3$\%$ \\
Africa & 33.4 & 10.9 & 14.7 & 96.3$\%$ \\
Scotland & 23.0 & 1.5 & 22.8 & 9.4$\%$ \\
Germany & 95.9 & 0.9 & 16.7 & 15.9$\%$ \\
Philippines & 7.7 & 2.7 & 9.6 & 60.6$\%$ \\
Ireland & 23.8 & 0.8 & 7.2 & 20.4$\%$ \\
Malaysia & 2.2 & 0.1 & 1.4 & 7.1$\%$ \\
\hline
\end{tabular}
\label{data}
\end{center}
\end{table}

\begin{table*}[htb]
\caption{Number of distinct speakers in CommonVoice for various accents}
\begin{center}
\begin{tabular}{lrrrrrrrrrrrr}
\hline
Accent & England & US & India & Germany & Africa & Canada & Australia & Philippines & Scotland & Ireland & Malaysia & Wales \\
\hline
Filtered   & 583  & 4335  & 1171  & \textbf{1}    & 71   & 599   & 493   & 55    & 73   & 17   & \textbf{10}   & \textbf{4}   \\
Unfiltered & 1507 & 5218  & 1254  & \textbf{3}    & 178  & 635   & 540   & 83    & 109  & 123  & 60   & 49  \\
Unlabeled  & 3580 & 21276 & 7124  & 2458 & 586  & 26459 & 5093  & 1315  & 3179 & 1392 & 391  & 449 \\
\hline
\end{tabular}
\end{center}
\label{speakers}
\end{table*}


\begin{table}[htbp]
\caption{Precision on CommonAccent test}
\begin{center}
\begin{tabular}{|c|c|c|c|}
\hline
\textbf{Accent} & \textbf{Geolocation} & \textbf{XLSR} & \textbf{GenAID} \\
\hline
US & 78$\%$ & 86$\%$ & \textbf{89\%} \\
England & 66$\%$ & 67$\%$ & \textbf{87\%} \\
India & \textbf{87\%} & 61$\%$ & \textbf{87\%} \\
Canada & 9$\%$ & 11$\%$ & \textbf{21\%} \\
Australia & \textbf{70\%} & 47$\%$ & \textbf{70\%} \\
Africa & \textbf{95\%} & 16$\%$ & $74\%$ \\
Scotland & 17$\%$ & 23$\%$ & \textbf{77\%} \\
Philippines & \textbf{77\%} & 5$\%$ & $46\%$ \\
Ireland & 3$\%$ & 11$\%$ & \textbf{62\%} \\
\hline
\end{tabular}
\label{precision}
\end{center}
\end{table}

Table~\ref{data} shows the amount of self-reported accent data pre and post filtering, as well as the amount of available data when the geolocation model~\cite{geolocating} is directly employed for accent label discovery. We additionally compute the precision of label discovery using the geolocation model on all the accents included in the CommonAccent dataset, as shown in table \ref{precision}. We observe that our method outperforms XLSR~\cite{xlsr} fine-tuned on CommonAccent, and is comparable with GenAID~\cite{accentbox} on many accents, in terms of label discovery precision, the most important metric for ensuring that the discovered accent data is of high quality.

The geolocation method for accent label discovery may theoretically be extended to any accents that are associated with a geographical range. In practice, since the model was trained on radio broadcasts, its performance depends on the availability and location of broadcasts spoken in the desired accent, as well as which broadcasts were chosen in training. The model works well for Indian and Australian accents, but somewhat worse for accents of the British Isles. 

\subsection{Fine-tuning XTTS-v2 for Accented TTS}

XTTS-v2~\cite{xtts} is a multilingual TTS model trained on CommonVoice. It consists of three main components: a discrete VAE~\cite{dvae} encoder, a discrete token sequence to sequence transformer~\cite{transformer}, and a Hifi-GAN~\cite{hifigan} vocoder which converts token sequence predictions into waveforms. The language of synthesized speech is controlled with a language token that is pre-pended to each sequence. Notably, XTTS-v2 exhibits some zero-shot accented TTS capability when the language token used during inference is different from the language of the underlying text. 

We fine-tune our accented TTS model from XTTS-v2, replacing the language token with an accent token as we're limited to TTS in English for this study. During training, we up-sample lower-resource accents so that all the accents included in our training are equally represented in each batch.


\subsection{Timbre and Acoustic Diversification using kNN-VC}

In a low-resource accented TTS training, it is common for certain accents to have very few distinct speakers in the dataset, causing timbre generalization difficulties for the model on those accents. 
Table~\ref{speakers} shows that the CommonVoice dataset suffers from this very issue, with certain accents (German before filtering; German and Welsh after filtering) containing less than $10$ distinct speakers in their respective training set. 
Previous works~\cite{nansy, radmmm, speaker_augmentation, without_frontend} have employed various data augmentation techniques, including duration, pitch, phonetic feature perturbation, as well as voice conversion, in order to promote separation between different aspects of speech: timbre, accent and content.

In our work, we employ a voice conversion system known as kNN-VC~\cite{knn_vc} to augment our data. Due to the strict time alignment of kNN-VC's input and output speech, it is known to be able to faithfully produce speech with the timbre of the target speaker while preserving the input utterance's accent and speaking style~\cite{vpc}. We convert the timbre of each utterance to a randomly chosen speaker from LibriTTS~\cite{libritts}, allowing the model to be exposed to a variety of speaker timbres in each accent.

\section{Experiments}

\subsection{XTTS-v2 Fine-tuning}

We included all the accents in CommonVoice that has more than $2$ hours of training data (as shown in table~\ref{data}). We do not measure the multilingual TTS capability of the model post fine-tuning. We fine-tune a separate model using each of the data label filtering/discovery strategy discussed in section \ref{appproaches} for comparison.

We perform ablation and comparison studies for kNN-VC augmentation on the XTTS-v2 model fine-tuned on the filtered (Approach~\ref{filtered_approach}) set, 
training one model without any data augmentation
and another with pitchshift augmentation where the fractional step parameter is chosen at random between $-4$ and $4$.

\subsection{Baseline Systems}

In addition to the original XTTS-v2 system, we compare our fine-tuned models with two external baselines: AccentBox\footnote{https://github.com/jzmzhong/coqui-TTS/tree/accentbox/}~\cite{accentbox} and CosyVoice2\footnote{https://github.com/FunAudioLLM/CosyVoice2}~\cite{cosyvoice2}. 

\subsubsection{AccentBox}

Accentbox was pretrained on LibriTTS-R~\cite{libritts-r} and subsequently fine-tuned on VCTK~\cite{vctk}, which contains $11$ different accents from the British isles. As accented TTS inference for AccentBox is performed using an accent embedding as conditioning, we generate accented speech using AccentBox using the following procedure: first, for each accent, we extract and average the embeddings of each utterance with that accent label in the CommonVoice development split; next, at inference time, we choose the average embedding of the desired accent label as the accent conditioning embedding. 

\subsubsection{CosyVoice2}

While CosyVoice2 does not explicitly support accent-controlled TTS in English, it performs zero-shot style copying and demonstrates some degree of accent imitation capabilities. We approximate accented TTS using CosyVoice2 by randomly picking an utterance from CommonVoice development split with the desired accent label to serve as the style prompt. 

\subsubsection{XTTS-v2 baseline configurations}

We experiment with two baseline configurations of XTTS-v2. In the first configuration, we run inference with English as the language label, which tends to generate US or English-accented speech. In the second configuration (labeled "XTTS approx" in result tables), we leverage XTTS-v2's limited zero-shot L2 accent synthesis capability, and pick an "approximate target" language label corresponding to the desired accent label whenever applicable (for example, we use language label "Hindi" when generating Indian-accented speech).

\subsection{Model-based Evaluation of Accent Similarity}

Following~\cite{evaluation}, we identified a number of candidates for model-based evaluation of accent similarity: vowel formants (VF), phonetic posteriorgrams (PPG)\footnote{https://github.com/liusongxiang/ppg-vc/tree/main}~\cite{ppg}, Mel ceptral distortion (MCD), WavLM~\cite{wavlm} fine-tuned for speaker identity\footnote{https://huggingface.co/microsoft/wavlm-base-plus-sv}), XLSR~\cite{xlsr} fine-tuned on CommonAccent\footnote{https://huggingface.co/Jzuluaga/accent-id-commonaccent\_xlsr-en-english}~\cite{commonaccent}, and GenAID\footnote{https://github.com/jzmzhong/GenAID/tree/GenAID}. 

\begin{figure}[htbp]
\centerline{\includegraphics[width=0.35\textwidth]{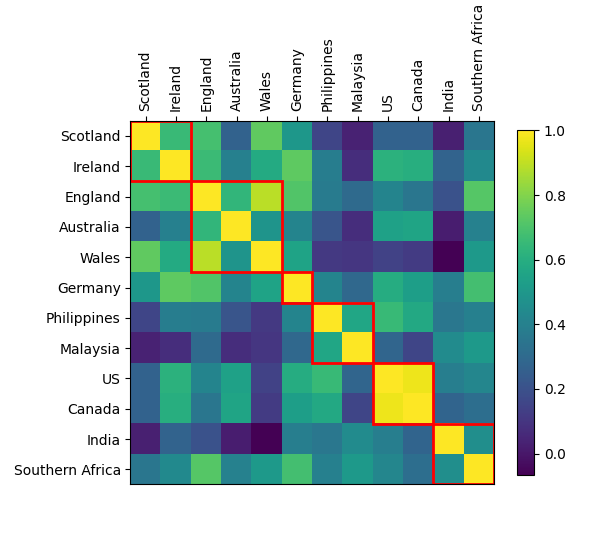}}
\caption{Cosine Similarity Matrix of average GenAID embeddings for various accents in CommonVoice.}
\label{fig:accent_similarity}
\vspace{-5mm}
\end{figure}

We first observed the distribution of similarity scores compute using each of these metrics on content-normalized but accent-diverse speech data from the Speech Accent Archive~\cite{speech_accent_archive}. We found no meaningful difference between the scores for different accents and those for the same accents in terms of VF, PPG, and MCD. The remaining three models all demonstrated the ability to distinguish speech accents in recorded speech; however, WavLM and XLSR were unable to meaningfully differentiate between synthesis systems, giving almost identical scores for each of our proposed and baseline systems. We therefore use GenAID as our primary model for accent evaluation. Figure~\ref{fig:accent_similarity}, where different accents were grouped using the spectral clustering algorithm ($n = 7$) according to their pairwise average GenAID embedding cosine similarity, illustrates that the cosine similarities align to an extent with established typological groupings of English accents.

We compute the following metrics using GenAID as our embedding extractor:

\begin{enumerate}[leftmargin=*, itemsep=0em]
    \item Table \ref{tab:genaid_gt_sim}: Accent embedding cosine similarity to the ground truth utterance \label{gt_sim};
    \item Table \ref{tab:auto_dcf}: Motivated by the  LRE22 challenge \cite{lre22}, we similarly construct an accent evaluation use GenAID as a ``frontend'' model. We use the evaluation data as enrollment. We train a Gaussian backend model on the ground-truth data, treating it as enrollment. We then use the synthetic speech as the evaluation data, where the goal is to produce accented enough speech that it can be easily detected. We measure the detection cost function (DCF) at two operating points, as done in \cite{lre22} to compute how close the generated speech was to the target accent. The goal is effectively to fool an accentID model. Due to the high dimensionality of GenAID embeddings and the relatively small sample size of our test set, we used a very reduced dimension of $18$ for PCA when training the Gaussian backend. \label{lre}
\end{enumerate}

Metric~\ref{gt_sim} is the most commonly used objective metric for accented TTS evaluation. However, it assumes that accent embeddings are disentangled from speaker or content information, a condition that is often not met in practice. 
Metric ~\ref{lre} addresses these limitations by aggregating the reference accent embeddings. In particular, metric~\ref{lre} accounts for different degrees of intra-accent embedding variance for each target accent. 
Furthermore, as metric ~\ref{lre} does not require synthesized utterances to have corresponding ground-truth utterances, we are able to greatly increase the test set (especially for low-resource accents) and control for factors such as speaker timbre and speech content when comparing across accents. 

\subsection{Human Subjective Evaluation}

For each target accent, we enlist $15$ human evaluators through prolific\footnote{https://app.prolific.com/} who are based in the country or region where the target accent would be prevalent or dominant. For each utterance, evaluators are tasked with producing two scores, both on a 1–5 scale: naturalness and accent plausibility. Following~\cite{eval}, we narrowly define naturalness as "sounding like it was produced by a real, human speaker rather than by a computer or artificial system". Accent plausibility is defined as "how closely the accent in the audio matches the way natives of [country/region] would naturally speak English." 
Each human evaluator is presented with $30$ real utterances from the CommonVoice test split with the target accent label, as well as 30 utterances from each system included in the human evaluation with the same content as the real utterances. An exception is made for Filipino-accented English, where, due to lack of data availability, only $20$ utterances from each system are included.

In this work, as we trained and evaluated on a much greater number of accents than is commonly reported, human evaluation of every system output for all the accents seen in training is prohibitively expensive. Furthermore, many accents share many similarities and appear to be quite confusable, i.e., US and Canadian English (see Figure \ref{fig:accent_similarity}). We therefore focused human evaluation on a subset of accents with wide geographic coverage, unique linguistic features or data particularities, and which facilitate comparison with prior work.

\section{Results}

\subsection{Speaker Similarity and Content Preservation}

We analyze the speaker timbre copying capability of the various models using Resemblyzer.\footnote{https://github.com/resemble-ai/Resemblyzer}
As shown in Table~\ref{speaker_sim}, our systems trained with data processed with numerous strategies (filtered, unfiltered and unlabeled) achieve very similar speaker similarity scores that are comparable with that of CosyVoice2. On the content preservation side, we measured the word error rate (WER) of synthesized speech using Whisper\footnote{https://github.com/openai/whisper}~\cite{whisper}. CosyVoice2 outperforms all other systems, although it is worth noting that synthesized accented speech presents a challenge to ASR systems.

\begin{table}[htbp]
\centering
\caption{Cosine similarity of Speaker Embedding and WER. Cells are colored based on their relative value in their respective column: darker indicates better performance.}
\begin{tabular}{l c c}
\toprule
\textbf{System} & \textbf{Speaker Sim. $\uparrow$} & \textbf{WER (\%) $\downarrow$} \\
\midrule
Ground Truth         &                                                     & \reversegradientcell{7.1}{7.1}{23.2}{white}{green}{50} \\
\midrule
Filtered             & \gradientcell{0.843}{0.66}{0.86}{white}{yellow}{50} & \reversegradientcell{13.1}{7.1}{23.2}{white}{green}{50}\\
Unfiltered           & \gradientcell{0.847}{0.66}{0.86}{white}{yellow}{50} & \reversegradientcell{14.0}{7.1}{23.2}{white}{green}{50}\\
Unlabeled            & \gradientcell{0.858}{0.66}{0.86}{white}{yellow}{50} & \reversegradientcell{18.2}{7.1}{23.2}{white}{green}{50}\\
\midrule
Accentbox            & \gradientcell{0.664}{0.66}{0.86}{white}{yellow}{50} & \reversegradientcell{14.2}{7.1}{23.2}{white}{green}{50}\\
CosyVoice2           & \gradientcell{0.855}{0.66}{0.86}{white}{yellow}{50} & \reversegradientcell{7.1}{7.1}{23.2}{white}{green}{50} \\
XTTS-v2              & \gradientcell{0.812}{0.66}{0.86}{white}{yellow}{50} & \reversegradientcell{8.6}{7.1}{23.2}{white}{green}{50} \\
XTTS approx          & \gradientcell{0.813}{0.66}{0.86}{white}{yellow}{50} & \reversegradientcell{23.2}{7.1}{23.2}{white}{green}{50}\\
\bottomrule
\label{speaker_sim}
\end{tabular}
\vspace{-5mm}
\end{table}

\subsection{Our system vs. Baselines}

\subsubsection{Objective Evaluations}
\begin{table*}[htb]
\centering
\caption{Automated evaluation based on GenAID embeddings. $^\dagger$ AccentBox was conditioned on the GenAID embeddings during training for accent generation, so this may not be a fair comparison. $^*$ The training data may not contain all accents, which could have contributed to poor performance on certain accents. XTTS-approx is the XTTS-v2 system where the English language token is replaced with a non-English L1 language that might more closely align with the accent, i.e., Hindi, for Indian accented English.}
\begin{subtable}{\textwidth}
\centering
\caption{ Cosine similarity of ground-truth utterances and generated accented speech samples prompted with different utterances from the same ground-truth speakers. Darker colors indicate better similarity.}
\adjustbox{max width=\linewidth}{
\begin{tabular}{l|
  C{0.75cm}| C{0.75cm} C{0.75cm} C{0.75cm} C{1cm} C{0.75cm} 
  C{0.75cm} C{1cm} C{1.2cm} C{1cm} C{0.75cm} C{1cm} C{0.75cm}}
\toprule
& \multicolumn{13}{c}{\textbf{Accent} (Cosine Similarity $\uparrow$)} \\
\multicolumn{1}{c|}{\textbf{System}} & All & England & US & India & Germany & Africa & Canada & Australia & Philippines & Scotland & Ireland & Malaysia & Wales \\
\midrule
XTTS-v2 & \gradientcell{0.356}{0.356}{0.542}{white}{blue}{50} & \gradientcell{0.358}{0.359}{0.510}{white}{blue}{50} & \gradientcell{0.437}{0.386}{0.560}{white}{blue}{50} & \gradientcell{0.270}{0.270}{0.629}{white}{blue}{50} & \gradientcell{0.382}{0.368}{0.470}{white}{blue}{50} & \gradientcell{0.302}{0.307}{0.466}{white}{blue}{50} & \gradientcell{0.412}{0.412}{0.603}{white}{blue}{50} & \gradientcell{0.348}{0.341}{0.509}{white}{blue}{50} & \gradientcell{0.304}{0.304}{0.568}{white}{blue}{50} & \gradientcell{0.348}{0.348}{0.546}{white}{blue}{50} & \gradientcell{0.371}{0.371}{0.451}{white}{blue}{50} & \gradientcell{0.245}{0.202}{0.593}{white}{blue}{50} & \gradientcell{0.401}{0.401}{0.561}{white}{blue}{50} \\

XTTS approx  & \gradientcell{0.411}{0.356}{0.542}{white}{blue}{50} & \gradientcell{0.364}{0.359}{0.510}{white}{blue}{50} & \gradientcell{0.440}{0.386}{0.560}{white}{blue}{50} & \gradientcell{0.459}{0.270}{0.629}{white}{blue}{50} & \gradientcell{0.470}{0.368}{0.470}{white}{blue}{50} & \gradientcell{0.374}{0.307}{0.466}{white}{blue}{50} & \gradientcell{0.427}{0.412}{0.603}{white}{blue}{50} & \gradientcell{0.341}{0.341}{0.509}{white}{blue}{50} & \gradientcell{0.475}{0.304}{0.568}{white}{blue}{50} & \gradientcell{0.361}{0.348}{0.546}{white}{blue}{50} & \gradientcell{0.370}{0.371}{0.451}{white}{blue}{50} & \gradientcell{0.496}{0.202}{0.593}{white}{blue}{50} & \gradientcell{0.411}{0.401}{0.561}{white}{blue}{50} \\
\midrule
Unlabeled & \gradientcell{0.485}{0.356}{0.542}{white}{blue}{50} & \gradientcell{0.396}{0.359}{0.510}{white}{blue}{50} & \gradientcell{0.489}{0.386}{0.560}{white}{blue}{50} & \gradientcell{0.606}{0.270}{0.629}{white}{blue}{50} & \gradientcell{0.368}{0.368}{0.470}{white}{blue}{50} & \gradientcell{0.392}{0.307}{0.466}{white}{blue}{50} & \gradientcell{0.524}{0.412}{0.603}{white}{blue}{50} & \gradientcell{0.469}{0.341}{0.509}{white}{blue}{50} & \gradientcell{0.510}{0.304}{0.568}{white}{blue}{50} & \gradientcell{0.480}{0.348}{0.546}{white}{blue}{50} & \gradientcell{0.391}{0.371}{0.451}{white}{blue}{50} & \gradientcell{0.407}{0.202}{0.593}{white}{blue}{50} & \gradientcell{0.487}{0.401}{0.561}{white}{blue}{50} \\

Unfiltered & \gradientcell{0.499}{0.356}{0.542}{white}{blue}{50} & \gradientcell{0.436}{0.359}{0.510}{white}{blue}{50} & \gradientcell{0.488}{0.386}{0.560}{white}{blue}{50} & \gradientcell{0.602}{0.270}{0.629}{white}{blue}{50} & \gradientcell{0.444}{0.368}{0.470}{white}{blue}{50} & \gradientcell{0.371}{0.307}{0.466}{white}{blue}{50} & \gradientcell{0.549}{0.412}{0.603}{white}{blue}{50} & \gradientcell{0.471}{0.341}{0.509}{white}{blue}{50} & \gradientcell{0.534}{0.304}{0.568}{white}{blue}{50} & \gradientcell{0.485}{0.348}{0.546}{white}{blue}{50} & \gradientcell{0.447}{0.371}{0.451}{white}{blue}{50} & \gradientcell{0.522}{0.202}{0.593}{white}{blue}{50} & \gradientcell{0.518}{0.401}{0.561}{white}{blue}{50} \\

Filtered & \gradientcell{0.520}{0.356}{0.542}{white}{blue}{50} & \gradientcell{0.451}{0.359}{0.510}{white}{blue}{50} & \gradientcell{0.507}{0.386}{0.560}{white}{blue}{50} & \gradientcell{0.629}{0.270}{0.629}{white}{blue}{50} & \gradientcell{0.409}{0.368}{0.470}{white}{blue}{50} & \gradientcell{0.396}{0.307}{0.466}{white}{blue}{50} & \gradientcell{0.579}{0.412}{0.603}{white}{blue}{50} & \gradientcell{0.489}{0.341}{0.509}{white}{blue}{50} & \gradientcell{0.568}{0.304}{0.568}{white}{blue}{50} & \gradientcell{0.546}{0.348}{0.546}{white}{blue}{50} & \gradientcell{0.451}{0.371}{0.451}{white}{blue}{50} & \gradientcell{0.593}{0.202}{0.593}{white}{blue}{50} & \gradientcell{0.527}{0.401}{0.561}{white}{blue}{50} \\
\midrule
AccentBox$^\dagger$$^*$ & \gradientcell{0.376}{0.356}{0.542}{white}{blue}{50} & \gradientcell{0.384}{0.359}{0.510}{white}{blue}{50} & \gradientcell{0.386}{0.386}{0.560}{white}{blue}{50} & \gradientcell{0.334}{0.270}{0.629}{white}{blue}{50} & \gradientcell{0.376}{0.368}{0.470}{white}{blue}{50} & \gradientcell{0.307}{0.307}{0.466}{white}{blue}{50} & \gradientcell{0.415}{0.412}{0.603}{white}{blue}{50} & \gradientcell{0.439}{0.341}{0.509}{white}{blue}{50} & \gradientcell{0.338}{0.304}{0.568}{white}{blue}{50} & \gradientcell{0.426}{0.348}{0.546}{white}{blue}{50} & \gradientcell{0.380}{0.371}{0.451}{white}{blue}{50} & \gradientcell{0.202}{0.202}{0.593}{white}{blue}{50} & \gradientcell{0.430}{0.401}{0.561}{white}{blue}{50} \\
CosyVoice2$^*$ & \gradientcell{0.542}{0.356}{0.542}{white}{blue}{50} & \gradientcell{0.510}{0.359}{0.510}{white}{blue}{50} & \gradientcell{0.560}{0.386}{0.560}{white}{blue}{50} & \gradientcell{0.597}{0.270}{0.629}{white}{blue}{50} & \gradientcell{0.428}{0.368}{0.470}{white}{blue}{50} & \gradientcell{0.466}{0.307}{0.466}{white}{blue}{50} & \gradientcell{0.603}{0.412}{0.603}{white}{blue}{50} & \gradientcell{0.509}{0.341}{0.509}{white}{blue}{50} & \gradientcell{0.566}{0.304}{0.568}{white}{blue}{50} & \gradientcell{0.439}{0.348}{0.546}{white}{blue}{50} & \gradientcell{0.432}{0.371}{0.451}{white}{blue}{50} & \gradientcell{0.521}{0.202}{0.593}{white}{blue}{50} & \gradientcell{0.561}{0.401}{0.561}{white}{blue}{50} \\
\bottomrule
\end{tabular}
}
\label{tab:genaid_gt_sim}
\end{subtable}

\vspace{1em}

\begin{subtable}{\textwidth}
\centering
\caption{The detection cost function (DCF) averaged over two operation points ($p_{tgt} = 0.1$ and $p_{tgt} = 0.5$). The DCF is computed using a Gaussian backend with shared covariance. The real, ground-truth evaluation data is used for enrollment, and the generated synthetic data are the test set.}
\adjustbox{max width=\linewidth}{
\begin{tabular}{l|
  C{0.75cm}| C{0.75cm} C{0.75cm} C{0.75cm} C{1cm} C{0.75cm} 
  C{0.75cm} C{1cm} C{1.2cm} C{1cm} C{0.75cm} C{1cm} C{0.75cm}}
\toprule
& \multicolumn{13}{c}{\textbf{Accent} (Detection Cost Function
$\downarrow$)} \\
\multicolumn{1}{c|}{\textbf{System}} & All & England & US & India & Germany & Africa & Canada & Australia & Philippines & Scotland & Ireland & Malaysia & Wales \\
\midrule
XTTS-v2 & \reversegradientcell{1.191}{0.571}{1.191}{white}{blue}{50} & \reversegradientcell{1.454}{0.652}{1.454}{white}{blue}{50} & \reversegradientcell{1.285}{0.674}{1.285}{white}{blue}{50} & \reversegradientcell{1.001}{0.494}{0.945}{white}{blue}{50} & \reversegradientcell{1.434}{0.570}{1.439}{white}{blue}{50} & \reversegradientcell{1.133}{0.861}{1.133}{white}{blue}{50} & \reversegradientcell{1.184}{0.672}{1.184}{white}{blue}{50} & \reversegradientcell{1.137}{0.382}{1.137}{white}{blue}{50} & \reversegradientcell{1.044}{0.489}{1.044}{white}{blue}{50} & \reversegradientcell{1.146}{0.442}{1.146}{white}{blue}{50} & \reversegradientcell{1.074}{0.683}{1.074}{white}{blue}{50} & \reversegradientcell{1.013}{0.319}{1.046}{white}{blue}{50} & \reversegradientcell{1.385}{0.511}{1.385}{white}{blue}{50} \\

XTTS approx & \reversegradientcell{1.136}{0.571}{1.191}{white}{blue}{50} 
& \reversegradientcell{1.199}{0.652}{1.454}{white}{blue}{50} 
& \reversegradientcell{1.142}{0.674}{1.285}{white}{blue}{50} 
& \reversegradientcell{0.967}{0.494}{0.945}{white}{blue}{50} 
& \reversegradientcell{1.439}{0.570}{1.439}{white}{blue}{50} 
& \reversegradientcell{1.133}{0.861}{1.133}{white}{blue}{50} 
& \reversegradientcell{1.044}{0.672}{1.184}{white}{blue}{50} 
& \reversegradientcell{1.098}{0.382}{1.137}{white}{blue}{50} 
& \reversegradientcell{1.042}{0.489}{1.044}{white}{blue}{50} 
& \reversegradientcell{1.094}{0.442}{1.146}{white}{blue}{50} 
& \reversegradientcell{1.053}{0.683}{1.074}{white}{blue}{50} 
& \reversegradientcell{0.911}{0.319}{1.046}{white}{blue}{50} 
& \reversegradientcell{1.157}{0.511}{1.385}{white}{blue}{50} \\
\midrule
Unlabeled & \reversegradientcell{0.936}{0.571}{1.191}{white}{blue}{50} & \reversegradientcell{1.092}{0.652}{1.454}{white}{blue}{50} & \reversegradientcell{0.869}{0.674}{1.285}{white}{blue}{50} & \reversegradientcell{0.553}{0.494}{0.945}{white}{blue}{50} & \reversegradientcell{1.269}{0.570}{1.439}{white}{blue}{50} & \reversegradientcell{1.088}{0.861}{1.133}{white}{blue}{50} & \reversegradientcell{0.799}{0.672}{1.184}{white}{blue}{50} & \reversegradientcell{0.861}{0.382}{1.137}{white}{blue}{50} & \reversegradientcell{0.752}{0.489}{1.044}{white}{blue}{50} & \reversegradientcell{0.998}{0.442}{1.146}{white}{blue}{50} & \reversegradientcell{0.996}{0.683}{1.074}{white}{blue}{50} & \reversegradientcell{1.046}{0.319}{1.046}{white}{blue}{50} & \reversegradientcell{0.906}{0.511}{1.385}{white}{blue}{50} \\

Unfiltered & \reversegradientcell{0.735}{0.571}{1.191}{white}{blue}{50} & \reversegradientcell{0.730}{0.652}{1.454}{white}{blue}{50} & \reversegradientcell{0.795}{0.674}{1.285}{white}{blue}{50} & \reversegradientcell{0.651}{0.494}{0.945}{white}{blue}{50} & \reversegradientcell{0.756}{0.570}{1.439}{white}{blue}{50} & \reversegradientcell{0.933}{0.861}{1.133}{white}{blue}{50} & \reversegradientcell{0.748}{0.672}{1.184}{white}{blue}{50} & \reversegradientcell{0.580}{0.382}{1.137}{white}{blue}{50} & \reversegradientcell{0.750}{0.489}{1.044}{white}{blue}{50} & \reversegradientcell{0.764}{0.442}{1.146}{white}{blue}{50} & \reversegradientcell{0.725}{0.683}{1.074}{white}{blue}{50} & \reversegradientcell{0.788}{0.319}{1.046}{white}{blue}{50} & \reversegradientcell{0.600}{0.511}{1.385}{white}{blue}{50} \\

Filtered & \reversegradientcell{0.571}{0.571}{1.191}{white}{blue}{50} & \reversegradientcell{0.652}{0.652}{1.454}{white}{blue}{50} & \reversegradientcell{0.674}{0.674}{1.285}{white}{blue}{50} & \reversegradientcell{0.494}{0.494}{0.945}{white}{blue}{50} & \reversegradientcell{0.638}{0.570}{1.439}{white}{blue}{50} & \reversegradientcell{0.869}{0.861}{1.133}{white}{blue}{50} & \reversegradientcell{0.672}{0.672}{1.184}{white}{blue}{50} & \reversegradientcell{0.382}{0.382}{1.137}{white}{blue}{50} & \reversegradientcell{0.489}{0.489}{1.044}{white}{blue}{50} & \reversegradientcell{0.442}{0.442}{1.146}{white}{blue}{50} & \reversegradientcell{0.701}{0.683}{1.074}{white}{blue}{50} & \reversegradientcell{0.329}{0.319}{1.046}{white}{blue}{50} & \reversegradientcell{0.511}{0.511}{1.385}{white}{blue}{50} \\
\midrule
AccentBox$^\dagger$$^*$ & \reversegradientcell{0.862}{0.571}{1.191}{white}{blue}{50} & \reversegradientcell{0.926}{0.652}{1.454}{white}{blue}{50} & \reversegradientcell{0.771}{0.674}{1.285}{white}{blue}{50} & \reversegradientcell{0.945}{0.494}{0.945}{white}{blue}{50} & \reversegradientcell{1.227}{0.570}{1.439}{white}{blue}{50} & \reversegradientcell{0.920}{0.861}{1.133}{white}{blue}{50} & \reversegradientcell{0.784}{0.672}{1.184}{white}{blue}{50} & \reversegradientcell{0.855}{0.382}{1.137}{white}{blue}{50} & \reversegradientcell{0.902}{0.489}{1.044}{white}{blue}{50} & \reversegradientcell{0.449}{0.442}{1.146}{white}{blue}{50} & \reversegradientcell{0.683}{0.683}{1.074}{white}{blue}{50} & \reversegradientcell{1.001}{0.319}{1.046}{white}{blue}{50} & \reversegradientcell{0.882}{0.511}{1.385}{white}{blue}{50} \\

CosyVoice2$^*$ & \reversegradientcell{0.748}{0.571}{1.191}{white}{blue}{50} & \reversegradientcell{0.963}{0.652}{1.454}{white}{blue}{50} & \reversegradientcell{0.937}{0.674}{1.285}{white}{blue}{50} & \reversegradientcell{0.539}{0.494}{0.945}{white}{blue}{50} & \reversegradientcell{0.570}{0.570}{1.439}{white}{blue}{50} & \reversegradientcell{0.861}{0.861}{1.133}{white}{blue}{50} & \reversegradientcell{0.847}{0.672}{1.184}{white}{blue}{50} & \reversegradientcell{0.604}{0.382}{1.137}{white}{blue}{50} & \reversegradientcell{0.526}{0.489}{1.044}{white}{blue}{50} & \reversegradientcell{0.683}{0.442}{1.146}{white}{blue}{50} & \reversegradientcell{0.891}{0.683}{1.074}{white}{blue}{50} & \reversegradientcell{0.600}{0.319}{1.046}{white}{blue}{50} & \reversegradientcell{0.956}{0.511}{1.385}{white}{blue}{50} \\
\bottomrule
\end{tabular}
}
\label{tab:auto_dcf}
\end{subtable}
\end{table*}

As shown in table \ref{tab:genaid_gt_sim}, CosyVoice2 outperforms all of our systems on metric \ref{gt_sim}, while Accentbox lags substantially behind. However, different patterns emerge when the influence of speaker similarity is eliminated, either by aggregating the accent embedding similarity target
or using the entire pool of target accent embeddings as enrollment data (metric \ref{lre}). Shwon in table \ref{tab:auto_dcf}, in most categories, the filtered system stands out for producing the most accent-appropriate utterances according to GenAID embeddings. Accentbox, whose training data included only accents from the British Isles, nevertheless stands out in a number of L1 accents including English, Australian, Scottish and Irish. CosyVoice2 remains competitive for nearly every accent, including L2 accents such as Indian, German and Malaysian. Finally, the unaccented XTTS-v2 baselines are not competitive with other baselines and models, except in rare cases like zero-shot generation of German-accented speech.

\subsubsection{Human Evaluations}
\begin{table*}[htb]
\centering
\scriptsize
\caption{Performance against SOTA systems}
\begin{tabular}{l|cccc|cccc}
\toprule
\multirow{2}{*}{\textbf{System}} &  \multicolumn{4}{c|}{\underline{\textbf{NMOS $\pm95\%$ CI}}} &  \multicolumn{4}{c}{\underline{\textbf{AMOS $\pm95\%$ CI}}}\\
 & \textbf{Ireland} & \textbf{India} & \textbf{Australia} & \textbf{US} & \textbf{Ireland} & \textbf{India} & \textbf{Australia} & \textbf{US} \\
\midrule
Ground Truth
    & \gradientcell{3.66}{2}{3.75}{white}{red}{50}$\pm 0.11$
    & \gradientcell{3.44}{2}{3.75}{white}{red}{50}$\pm 0.11$
    & \gradientcell{3.04}{2}{3.75}{white}{red}{50}$\pm 0.12$
    & \gradientcell{3.59}{2}{3.75}{white}{red}{50}$\pm 0.12$
    & \gradientcell{3.49}{2}{3.75}{white}{teal}{50}$\pm 0.13$
    & \gradientcell{3.33}{2}{3.75}{white}{teal}{50}$\pm 0.11$
    & \gradientcell{3.04}{2}{3.75}{white}{teal}{50}$\pm 0.12$
    & \gradientcell{3.43}{2}{3.75}{white}{teal}{50}$\pm 0.12$\\
\midrule
CosyVoice2
    & \gradientcell{2.92}{2}{3.75}{white}{red}{50}$\pm 0.11$
    & \gradientcell{3.30}{2}{3.75}{white}{red}{50}$\pm 0.11$
    & \gradientcell{3.18}{2}{3.75}{white}{red}{50}$\pm 0.12$
    & \gradientcell{3.70}{2}{3.75}{white}{red}{50}$\pm 0.11$
    & \gradientcell{2.09}{2}{3.75}{white}{teal}{50}$\pm 0.12$
    & \gradientcell{3.01}{2}{3.75}{white}{teal}{50}$\pm 0.12$
    & \gradientcell{2.95}{2}{3.75}{white}{teal}{50}$\pm 0.13$
    & \gradientcell{3.70}{2}{3.75}{white}{teal}{50}$\pm 0.11$\\
Accentbox 
    & \gradientcell{2.77}{2}{3.75}{white}{red}{50}$\pm 0.11$
    & 
    & 
    & \gradientcell{3.69}{2}{3.75}{white}{red}{50}$\pm 0.11$
    & \gradientcell{3.03}{2}{3.75}{white}{teal}{50}$\pm 0.12$
    & 
    & 
    & \gradientcell{3.63}{2}{3.75}{white}{teal}{50}$\pm 0.11$\\
\midrule
Filtered
    & \gradientcell{2.96}{2}{3.75}{white}{red}{50}$\pm 0.12$
    & \gradientcell{3.23}{2}{3.75}{white}{red}{50}$\pm 0.12$
    & \gradientcell{2.95}{2}{3.75}{white}{red}{50}$\pm 0.11$
    & \gradientcell{3.34}{2}{3.75}{white}{red}{50}$\pm 0.12$
    & \gradientcell{2.63}{2}{3.75}{white}{teal}{50}$\pm 0.12$
    & \gradientcell{3.04}{2}{3.75}{white}{teal}{50}$\pm 0.12$
    & \gradientcell{3.13}{2}{3.75}{white}{teal}{50}$\pm 0.12$
    & \gradientcell{3.42}{2}{3.75}{white}{teal}{50}$\pm 0.12$\\
\bottomrule
\end{tabular}
\label{tab:mos_baseline}
\end{table*}

CosyVoice2 was included in $4$ sets of tests: Irish, Indian, Australian, and US accents. 
Annotators consistently placed CosyVoice2 above every other system included in this study, even preferring it over real utterances in the US accent test (more discussions in section \ref{human_discussions}). CosyVoice2 also scored well on accent plausibility in every test except in the case of Irish.

Accentbox was included in $2$ sets of tests: Irish and US. Despite only being trained on accents from the British Isles, the original authors reported their capability to generate the unseen US accent. This was confirmed in our study, which found that annotators consistently preferred Accentbox over every other system in both tests where it was included. In the case of Irish accent, this came at the cost of slightly reduced naturalness.

The original XTTS-v2 was included in evaluations for Australian, Filipino, and US accents. As XTTS-v2 produces mostly acoustically clean, US or English-accented speech, it unsurprisingly scored very well on naturalness on all tests as well as accent plausibility on the US accent test. It scored poorly on accent plausibility in the Australian and Filipino accent evaluations.

\subsection{Filtering vs. No Filtering vs. Label Discovery}

\subsubsection{Objective Evaluations}
The model trained with data filtered using the geolocation model outperformed the model trained with self-reported accent labels in CommonVoice on nearly every accent. Some notable exceptions include German and Irish. Interestingly, contrary to our expectations, the filtering method resulted in smaller improvements in accents where we expect the labels to be noisier, namely English and US accents. 

According to both metrics, the model trained using labels discovered with the geolocation model underperformed the other data settings, unfiltered and filtered. However, it substantially outperformed the XTTS-v2 baselines on most accents. 

\subsubsection{Human Evaluations}
\begin{table*}[htb]
\centering
\caption{Human Evaluation: Comparison between different training data filtering/discovery strategies}
\begin{tabular}{l|ccc|ccc}
\toprule
\multirow{2}{*}{\textbf{System}} &  \multicolumn{3}{c|}{\underline{\textbf{NMOS $\pm95\%$ CI}}} &  \multicolumn{3}{c}{\underline{\textbf{AMOS $\pm95\%$ CI}}}\\
 & \textbf{Australia} & \textbf{Philippines} & \textbf{US} & \textbf{Australia} & \textbf{Philippines} & \textbf{US}  \\
\midrule
XTTS-v2
    & \gradientcell{2.85}{2}{3.75}{white}{red}{50}$\pm 0.11$
    & \gradientcell{2.85}{2}{3.75}{white}{red}{50}$\pm 0.19$
    & \gradientcell{3.45}{2}{3.75}{white}{red}{50}$\pm 0.12$
    & \gradientcell{2.22}{2}{3.75}{white}{teal}{50}$\pm 0.11$
    & \gradientcell{2.66}{2}{3.75}{white}{teal}{50}$\pm 0.19$
    & \gradientcell{3.61}{2}{3.75}{white}{teal}{50}$\pm 0.12$\\
\midrule
Unlabeled
    & \gradientcell{2.58}{2}{3.75}{white}{red}{50}$\pm 0.11$
    & \gradientcell{2.56}{2}{3.75}{white}{red}{50}$\pm 0.19$
    & \gradientcell{3.29}{2}{3.75}{white}{red}{50}$\pm 0.12$
    & \gradientcell{2.81}{2}{3.75}{white}{teal}{50}$\pm 0.12$
    & \gradientcell{2.52}{2}{3.75}{white}{teal}{50}$\pm 0.18$
    & \gradientcell{3.37}{2}{3.75}{white}{teal}{50}$\pm 0.12$\\
Unfiltered
    & \gradientcell{3.04}{2}{3.75}{white}{red}{50}$\pm 0.10$
    & \gradientcell{2.84}{2}{3.75}{white}{red}{50}$\pm 0.19$
    & \gradientcell{3.33}{2}{3.75}{white}{red}{50}$\pm 0.12$
    & \gradientcell{2.92}{2}{3.75}{white}{teal}{50}$\pm 0.11$
    & \gradientcell{2.77}{2}{3.75}{white}{teal}{50}$\pm 0.18$
    & \gradientcell{3.41}{2}{3.75}{white}{teal}{50}$\pm 0.11$\\
Filtered
    & \gradientcell{2.95}{2}{3.75}{white}{red}{50}$\pm 0.11$
    & \gradientcell{3.17}{2}{3.75}{white}{red}{50}$\pm 0.19$
    & \gradientcell{3.34}{2}{3.75}{white}{red}{50}$\pm 0.12$
    & \gradientcell{3.13}{2}{3.75}{white}{teal}{50}$\pm 0.12$
    & \gradientcell{2.98}{2}{3.75}{white}{teal}{50}$\pm 0.18$
    & \gradientcell{3.42}{2}{3.75}{white}{teal}{50}$\pm 0.12$\\
\midrule
Ground Truth
    & \gradientcell{3.04}{2}{3.75}{white}{red}{50}$\pm 0.12$
    & \gradientcell{3.68}{2}{3.75}{white}{red}{50}$\pm 0.18$
    & \gradientcell{3.59}{2}{3.75}{white}{red}{50}$\pm 0.12$
    & \gradientcell{3.04}{2}{3.75}{white}{teal}{50}$\pm 0.12$
    & \gradientcell{3.57}{2}{3.75}{white}{teal}{50}$\pm 0.17$
    & \gradientcell{3.43}{2}{3.75}{white}{teal}{50}$\pm 0.12$\\
\bottomrule
\end{tabular}
\label{tab:mos_label}
\end{table*}

Comparisons of human evaluation of the different filtering systems are shown in table \ref{tab:mos_label}. The unfiltered system, trained using self-reported accent labels from CommonVoice, was included in evaluations for Australian, Filipino, and US accents. The filtered system outperformed the unfiltered system substantially in both the Australian and the Filipino accents, demonstrating that data filtering using the geolocation model is effective at improving accented speech synthesis.

When it comes to the system trained with geolocation-discovered labels, results demonstrate decent accent generation capability of the resulting model, although annotators consistently prefer other systems over the unlabeled one in terms of accent plausibility. 

\subsection{Effects of kNN-VC Data Augmentation}

\subsubsection{Objective Evaluations}
\begin{table*}[htb]
\centering
\caption{Automated evaluation using the detection cost function (DCF) as described in Table \ref{tab:auto_dcf} on models trained using different data augmentation methods.}
\adjustbox{max width=\linewidth}{
\label{knn_lid}
\begin{tabular}{l|
  C{0.75cm}| C{0.75cm} C{0.75cm} C{0.75cm} C{1cm} C{0.75cm} 
  C{0.75cm} C{1cm} C{1.2cm} C{1cm} C{0.75cm} C{1cm} C{0.75cm}}
\toprule
 & \multicolumn{13}{c}{\textbf{Accent} (Detection Cost Function
$\downarrow$)} \\
\textbf{Method} & All & England & US & India & Germany & Africa & Canada & Australia & Philippines & Scotland & Ireland & Malaysia & Wales \\
\midrule
None & 
\reversegradientcell{0.744}{0.571}{0.744}{white}{violet}{50} & 
\reversegradientcell{0.709}{0.652}{0.715}{white}{violet}{50} & 
\reversegradientcell{0.690}{0.674}{0.699}{white}{violet}{50} & 
\reversegradientcell{0.587}{0.494}{0.587}{white}{violet}{50} & 
\reversegradientcell{0.864}{0.638}{0.864}{white}{violet}{50} & 
\reversegradientcell{0.989}{0.869}{1.025}{white}{violet}{50} & 
\reversegradientcell{0.690}{0.672}{0.704}{white}{violet}{50} & 
\reversegradientcell{0.489}{0.382}{0.489}{white}{violet}{50} & 
\reversegradientcell{0.678}{0.489}{0.678}{white}{violet}{50} & 
\reversegradientcell{0.612}{0.442}{0.625}{white}{violet}{50} & 
\reversegradientcell{0.905}{0.701}{0.905}{white}{violet}{50} & 
\reversegradientcell{0.884}{0.319}{0.884}{white}{violet}{50} & 
\reversegradientcell{0.827}{0.511}{0.827}{white}{violet}{50} \\
Pitchshift & 
\reversegradientcell{0.666}{0.571}{0.744}{white}{violet}{50} & 
\reversegradientcell{0.715}{0.652}{0.715}{white}{violet}{50} & 
\reversegradientcell{0.699}{0.674}{0.699}{white}{violet}{50} & 
\reversegradientcell{0.576}{0.494}{0.587}{white}{violet}{50} & 
\reversegradientcell{0.726}{0.638}{0.864}{white}{violet}{50} & 
\reversegradientcell{1.025}{0.869}{1.025}{white}{violet}{50} & 
\reversegradientcell{0.704}{0.672}{0.704}{white}{violet}{50} & 
\reversegradientcell{0.445}{0.382}{0.489}{white}{violet}{50} & 
\reversegradientcell{0.575}{0.489}{0.678}{white}{violet}{50} & 
\reversegradientcell{0.625}{0.442}{0.625}{white}{violet}{50} & 
\reversegradientcell{0.845}{0.701}{0.905}{white}{violet}{50} & 
\reversegradientcell{0.319}{0.319}{0.884}{white}{violet}{50} & 
\reversegradientcell{0.744}{0.511}{0.827}{white}{violet}{50} \\
kNN-VC & 
\reversegradientcell{0.571}{0.571}{0.744}{white}{violet}{50} & 
\reversegradientcell{0.652}{0.652}{0.715}{white}{violet}{50} & 
\reversegradientcell{0.674}{0.674}{0.699}{white}{violet}{50} & 
\reversegradientcell{0.494}{0.494}{0.587}{white}{violet}{50} & 
\reversegradientcell{0.638}{0.638}{0.864}{white}{violet}{50} & 
\reversegradientcell{0.869}{0.869}{1.025}{white}{violet}{50} & 
\reversegradientcell{0.672}{0.672}{0.704}{white}{violet}{50} & 
\reversegradientcell{0.382}{0.382}{0.489}{white}{violet}{50} & 
\reversegradientcell{0.489}{0.489}{0.678}{white}{violet}{50} & 
\reversegradientcell{0.442}{0.442}{0.625}{white}{violet}{50} & 
\reversegradientcell{0.701}{0.701}{0.905}{white}{violet}{50} & 
\reversegradientcell{0.329}{0.319}{0.884}{white}{violet}{50} & 
\reversegradientcell{0.511}{0.511}{0.827}{white}{violet}{50} \\
\bottomrule
\end{tabular}
}
\end{table*}

To study the effect of kNN-VC data augmentation, we perform two ablation studies: one with no data augmentation during training, and another using a signal-processing-based pitch augmentation method known as pitchshift as timbre augmentation during training. Table \ref{knn_lid} shows that on most accents, training with kNN-VC augmentation improves the accent similarity of the resulting model according to objective metrics. It is worth mentioning that kNN-VC introduces a significant training overhead, and that if training is time-constrained, then training with simple augmentation methods like pitchshift can also effectively improve the accent similarity of the model.

\subsubsection{Human Evaluations}
\begin{table*}[!htb]
\centering
\scriptsize
\caption{Human Evaluation: Effect of data augmentation during training}
\begin{tabular}{l|ccc|ccc}
\toprule
\multirow{2}{*}{\textbf{System}} &  \multicolumn{3}{c|}{\underline{\textbf{NMOS $\pm95\%$ CI}}} &  \multicolumn{3}{c}{\underline{\textbf{AMOS $\pm95\%$ CI}}}\\
 & \textbf{Australia} & \textbf{Scotland} & \textbf{US} & \textbf{Australia} & \textbf{Scotland} & \textbf{US}  \\
\midrule
Filtered - No Aug.
    & \gradientcell{2.81}{2}{3.75}{white}{red}{50}$\pm 0.11$
    & \gradientcell{3.33}{2}{3.75}{white}{red}{50}$\pm 0.11$
    & \gradientcell{3.40}{2}{3.75}{white}{red}{50}$\pm 0.11$
    & \gradientcell{2.90}{2}{3.75}{white}{teal}{50}$\pm 0.11$
    & \gradientcell{3.37}{2}{3.75}{white}{teal}{50}$\pm 0.11$
    & \gradientcell{3.55}{2}{3.75}{white}{teal}{50}$\pm 0.11$\\
Filtered - Pitchshift Aug.
    & \gradientcell{3.04}{2}{3.75}{white}{red}{50}$\pm 0.11$
    & \gradientcell{3.09}{2}{3.75}{white}{red}{50}$\pm 0.12$
    & \gradientcell{3.43}{2}{3.75}{white}{red}{50}$\pm 0.11$
    & \gradientcell{3.13}{2}{3.75}{white}{teal}{50}$\pm 0.11$
    & \gradientcell{3.16}{2}{3.75}{white}{teal}{50}$\pm 0.12$
    & \gradientcell{3.55}{2}{3.75}{white}{teal}{50}$\pm 0.11$\\
Filtered - kNN-VC aug.
    & \gradientcell{2.95}{2}{3.75}{white}{red}{50}$\pm 0.11$
    & \gradientcell{3.19}{2}{3.75}{white}{red}{50}$\pm 0.11$
    & \gradientcell{3.34}{2}{3.75}{white}{red}{50}$\pm 0.12$
    & \gradientcell{3.13}{2}{3.75}{white}{teal}{50}$\pm 0.12$
    & \gradientcell{3.17}{2}{3.75}{white}{teal}{50}$\pm 0.11$
    & \gradientcell{3.42}{2}{3.75}{white}{teal}{50}$\pm 0.12$\\
\midrule
Ground Truth
    & \gradientcell{3.04}{2}{3.75}{white}{red}{50}$\pm 0.12$
    & \gradientcell{3.63}{2}{3.75}{white}{red}{50}$\pm 0.11$
    & \gradientcell{3.59}{2}{3.75}{white}{red}{50}$\pm 0.12$
    & \gradientcell{3.04}{2}{3.75}{white}{teal}{50}$\pm 0.12$
    & \gradientcell{3.54}{2}{3.75}{white}{teal}{50}$\pm 0.12$
    & \gradientcell{3.43}{2}{3.75}{white}{teal}{50}$\pm 0.12$\\
\bottomrule
\end{tabular}
\label{tab:mos_aug}
\end{table*}

\setlength{\tabcolsep}{4pt}
\renewcommand{\arraystretch}{1.0}

\begin{table}[htb]
\centering
\caption{Human Evaluation: Effects of kNN-VC augmentation}
\vspace{-0.5em}
\adjustbox{max width=\linewidth}{
\begin{tabular}{@{}l|cc|cc@{}}
\toprule
\multirow{2}{*}{\textbf{Input}} &  \multicolumn{2}{c|}{\underline{\textbf{NMOS $\pm$95\% CI}}} &  \multicolumn{2}{c}{\underline{\textbf{AMOS $\pm$95\% CI}}} \\
 & \textbf{Australia} & \textbf{Philippines} & \textbf{Australia} & \textbf{Philippines} \\
\midrule
kNN-VC
    & \gradientcell{2.4}{2}{3.75}{white}{red}{50}$\pm 0.11$
    & \gradientcell{3.17}{2}{3.75}{white}{red}{50}$\pm 0.19$
    & \gradientcell{2.47}{2}{3.75}{white}{teal}{50}$\pm 0.12$
    & \gradientcell{3.14}{2}{3.75}{white}{teal}{50}$\pm 0.11$ \\
Ground Truth
    & \gradientcell{3.04}{2}{3.75}{white}{red}{50}$\pm 0.12$
    & \gradientcell{3.68}{2}{3.75}{white}{red}{50}$\pm 0.18$
    & \gradientcell{3.59}{2}{3.75}{white}{teal}{50}$\pm 0.12$
    & \gradientcell{3.57}{2}{3.75}{white}{teal}{50}$\pm 0.12$ \\
\bottomrule
\end{tabular}
}
\vspace{-0.5em}
\label{tab:mos_knnvc}
\end{table}

We first test the accent preservation effects of kNN-VC conversion. Real utterances voice converted using kNN-VC (into randomly-selected target speakers from LibriTTS) were included in two human evaluations: Filipino and Australian accents, as shown in table \ref{tab:mos_knnvc}. In the Filipino accent human evaluation, kNN-VC achieved by far the second best accent plausibility score, second only to real utterances. In the Australian accent human evaluation, kNN-VC received very poor scores on both naturalness (last) and accent plausibility (second from last, only above original XTTS-v2). We will further discuss this result in section \ref{human_discussions}.

Next, we compare the systems trained using different augmentation strategies. Ablation models - ones trained with no speaker augmentation and with pitchshift augmentation - were included in the studies for Australian, Scottish, and US accents. Human evaluators assessed these systems to be very similar in all three cases.

\subsection{Limitations of Human Evaluation} \label{human_discussions}

We observe that human annotators never give significantly higher accent plausibility scores than naturalness scores. Moreover, naturalness scores by human evaluators are known to be sensitive to acoustic conditions and content. These appear to be in effect in the US-accented human evaluation, where the relative noisiness of real utterance in CommonVoice resulted in many annotators preferring systems that generate acoustically clean utterances over real utterances. Similar effects likely contributed to the results in the Australian-accented evaluation, where both real utterances and kNN-VC voice-converted real utterances received relatively poor naturalness and accent plausibility scores.

\section{Conclusions}

In this work, we apply a geolocation model for accent label discovery and cleaning, a technique that can be extended to any quantity of unlabeled accented speech data as well as to previously unseen accent labels. We achieve precision comparable to SOTA baselines for accent ID. We subsequently fine-tune XTTS-v2 on CommonVoice data that had been accent-labeled or filtered using the geolocation method. To promote speaker-accent disentanglement, we apply kNN-VC augmentation to diversify the timbre and acoustics of the training data. Using these methods, we achieve competitive accented TTS results to existing approaches. 

Our future work will explore additional label discovery methods, such as pseudo-labeling using accent ID models, 
and will aim to bridge the gap between model-based and human evaluation of accented speech generation.

\section*{Acknowledgment}
This work was supported by the Office of the Director of National Intelligence (ODNI), Intelligence Advanced Research Projects Activity (IARPA), via the ARTS Program under contract D2023-2308110001. The views and conclusions contained herein are those of the authors and should not be interpreted as necessarily representing the official policies, either expressed or implied, of ODNI, IARPA, or the U.S. Government. The U.S. Government is authorized to reproduce and distribute reprints for governmental purposes notwithstanding any copyright annotation therein.

We would also like to mention generous help from Jan Melechovsky and Jinzuomu Zhong with setting up their models as baselines, Niyati Bafna on setting up accent ID baselines, and Lin Zhang for help with drafting the paper.
\bibliographystyle{IEEEtran}
\bibliography{mybib}

\end{document}